# Confirmation Bias in Generative AI Chatbots: Mechanisms, Risks, Mitigation Strategies, and Future Research Directions

Yiran Du[1]

[1]Institute of Cognitive Neuroscience, University College London, London, UK

**Abstract**
This article explores the phenomenon of confirmation bias in generative AI chatbots, a relatively underexamined aspect of AI-human interaction. Drawing on cognitive psychology and computational linguistics, it examines how confirmation bias—commonly understood as the tendency to seek information that aligns with existing beliefs—can be replicated and amplified by the design and functioning of large language models. The article analyzes the mechanisms by which confirmation bias may manifest in chatbot interactions, assesses the ethical and practical risks associated with such bias, and proposes a range of mitigation strategies. These include technical interventions, interface redesign, and policy measures aimed at promoting balanced AI-generated discourse. The article concludes by outlining future research directions, emphasizing the need for interdisciplinary collaboration and empirical evaluation to better understand and address confirmation bias in generative AI systems.

**Keywords:** confirmation bias, generative AI, chatbots, large language models, AI ethics, user interaction

## 1. Introduction

The emergence of generative AI chatbots has marked a significant turning point in the field of artificial intelligence (AI) (Chang et al., 2024). These systems, underpinned by large-scale language models, have demonstrated a remarkable capacity for producing coherent, contextually relevant, and often creative responses to human queries (Wang et al., 2024). Some of these models have been adapted to a wide range of applications, ranging from customer service to language translation, content generation, coding assistance, and various other domains where the capacity to generate natural-sounding text is valuable (Xi et al., 2025). While the impressive capabilities of generative AI chatbots have gained widespread attention, and their benefits are frequently highlighted, there are growing concerns regarding the unintended consequences that can arise from their deployment (Myers et al., 2024). One of these concerns revolves around the phenomenon of confirmation bias, which could manifest in AI-generated interactions in ways that have not yet been comprehensively analyzed by researchers.

Confirmation bias is well-studied in human cognition. It refers to the propensity of individuals to seek, interpret, and recall information that aligns with previously held beliefs, expectations, or assumptions (Klayman, 1995). Studies of human behavior have demonstrated that individuals often selectively attend to evidence that confirms their prior views, while discounting or overlooking contradictory data (Chester, 2021). This tendency can influence decision-making processes, interpersonal judgments, and the interpretation of data across a wide range of contexts (Del Vicario et al., 2017). In the realm of AI, discussions on bias frequently focus on issues such as racial bias, gender bias, or other forms of prejudiced outcomes in classification or recommendation systems (Chen et al., 2023). However, the more subtle phenomenon of confirmation bias in generative AI chatbots has received comparatively less attention. Despite this relative lack of focus, it might represent an important dimension of AI reliability and ethical deployment, as it can shape how users interact with these systems and how the systems, in turn, respond to users' prompts.

This perspectives article aims to provide a comprehensive discussion of confirmation bias in generative AI chatbots. It offers an overview of the concept of confirmation bias as understood in cognitive psychology, before examining how this phenomenon might manifest in the particular setting of AI-generated conversations. The mechanisms by which confirmation bias might arise in generative AI are explored, and the potential risks associated with this bias are analyzed from ethical, social, and practical standpoints. Strategies for mitigating confirmation bias and for designing chatbot systems that foster more balanced and reflective interactions are discussed, drawing upon insights from diverse fields such as user experience design, NLP interpretability research, and cognitive science. Finally, directions for future research are proposed, highlighting how interdisciplinary collaboration could further elucidate and address this phenomenon. By integrating perspectives from AI, psychology, and the humanities, it is hoped that a nuanced understanding of confirmation bias in generative AI chatbots will emerge, paving the way for the responsible and equitable use of these transformative technologies.

## 2. Conceptual Underpinnings of Confirmation Bias

Confirmation bias has been studied extensively in the domain of cognitive psychology. It emerged as a central concept in the 20th century, with researchers examining how humans handle information that aligns with or contradicts their existing beliefs (Wason, 1960, 1968). In a classic demonstration of confirmation bias, individuals presented with data sets frequently sought to confirm a favored hypothesis rather than looking for ways to refute it (Nickerson, 1998). This approach involves selective sampling of evidence, often driven by the need to reduce cognitive dissonance and maintain consistency in one's worldview (Lord et al., 1979). Over time, researchers uncovered a variety of manifestations of confirmation bias, including the bias in question framing, in the search for information, and in the interpretation of ambiguous data (Palminteri & Lebreton, 2022).

The theoretical foundation for confirmation bias draws from frameworks in cognitive psychology that emphasize heuristics and mental shortcuts, such as those first prominently identified in studies of bounded rationality (Rabin & Schrag, 1999). The assumption is that because of limited cognitive resources, individuals adopt strategies that minimize mental effort. Although such shortcuts can be adaptive in many contexts, they can also produce systematic errors in reasoning when individuals fail to update or revise prior beliefs in the face of new evidence (Nickerson, 1998). Confirmation bias, therefore, is often conceptualized as a byproduct of these cognitive constraints combined with motivational factors that shape how individuals navigate information (Wason, 1960). It has been argued that humans derive psychological comfort from confirming existing ideas, as challenges to these ideas can create a sense of unease (Wason, 1968).

Confirmation bias also intersects with social and cultural factors. In group settings, people often form echo chambers in which they reinforce each other's worldviews, thereby exacerbating the bias (Cookson et al., 2023). Social media platforms have been singled out for facilitating these echo chambers, as algorithms that recommend content relevant to a user's interests inadvertently expose them primarily to information that resonates with, rather than contradicts, their established beliefs (Alsaad et al., 2018). This can lead to polarization and tribalism, phenomena that have been studied with increased urgency in the last decade (Del Vicario et al., 2017). While confirmation bias is not solely responsible for such dynamics, it is often recognized as a powerful contributing force (Modgil et al., 2024).

It is therefore worthwhile to analyze whether generative AI chatbots, with their capacity to adapt to user queries, might inadvertently reproduce this bias. Users might bring preconceived beliefs or assumptions into their interactions with chatbots, and if the chatbot is shaped to provide contextually consistent or user-aligned responses, it could act as a digital echo chamber. The chatbot would then, effectively, confirm the user's perspective and fail to highlight any countervailing information. While humans can sometimes be alerted to the possibility of their biases through external interventions, the risk is that a generative AI chatbot, lacking built-in epistemological counter-checks, might not spontaneously introduce such interventions. Even if a user explicitly requests to be challenged or engaged in a debate, the chatbot's behavior might still be guided by patterns from its training data that favor simply continuing the user's line of reasoning, unless it is programmed to generate or present contradictory evidence.

It is important to recognize that confirmation bias in generative AI is not necessarily an outcome of malicious intent or an intrinsic design flaw. Rather, it might emerge from the probabilistic nature of how language models function (Chow et al., 2024). These models learn from enormous corpora of text, and their primary objective is to predict the next token or sequence of tokens (Nazi & Peng, 2024). When a user provides a prompt that contains a particular slant or assumption, the model calculates which subsequent words or phrases are most likely in the context of that prompt (Xi et al., 2025). It has been suggested that this process could produce a stable reinforcement of the user's assumption, especially in the absence of an explicit mechanism for contradictory or alternative discourse. As a result, the system's default behavior might be to comply with the user's viewpoint. Some generative AI systems even include design elements intended to be "helpful" and "friendly," potentially further inhibiting critical or adversarial responses to user prompts (Wang et al., 2024). This phenomenon becomes especially salient in scenarios where the user's prompt includes an implicit or explicit question about a controversial or polarizing topic.

The next section delves deeper into how confirmation bias may manifest in generative AI chatbots. This discussion draws upon observations from real-world deployments and from theoretical insights into AI development. Since this is a relatively new area of inquiry, the focus is on hypothesized and emergent patterns that might require further empirical validation. Nonetheless, by examining the potential channels through which confirmation bias can appear, it is hoped that a clear picture will emerge of the relevance of this bias to AI-generated text and the potential consequences for users and society at large.

## 3. Confirmation Bias in Generative AI Chatbots

Generative AI chatbots draw upon large language models that are often trained on massive, unfiltered datasets from the internet (Liang et al., 2024). These datasets encompass diverse textual materials, including news articles, social media posts, forum discussions, and other user-generated content. The fundamental mechanism behind generative models is to predict the next word in a sequence given the context of previous words. Through iterative training processes, the model learns statistical associations, contextual clues, and linguistic structures that can be used to generate coherent and contextually relevant text (Shanahan et al., 2023). The sophistication of these models has grown rapidly in recent years, with improvements in architecture, such as the shift from recurrent neural networks to transformers, along with exponential increases in model size, making generative text outputs more seamless and human-like (Wu et al., 2025).

While this technology offers numerous advantages, it also raises questions about how these systems handle queries that contain specific assumptions or biases. Generative AI models, by design, strive to produce responses that align with the user's prompt. If a user asks a question with a particular leaning, such as, "Why is the theory X definitely true?" the chatbot may begin its response with an assumption that theory X is indeed true. This dynamic is partly reflective of how human conversation typically unfolds, because human participants often adapt their speech to maintain coherence, politeness, or agreement. In the context of AI chatbots, however, this adaptation can become a computational manifestation of confirmation bias, as the system effectively confirms the user's premise rather than questioning it. When the conversation is extended over multiple turns, there can be further entrenchment, because the system's subsequent responses are increasingly grounded in the previously generated text, compounding the user's initial assumption.

There have been anecdotal accounts of generative chatbots providing elaborate justifications for erroneous or unfounded claims, apparently because the user's prompt presupposed those claims. In such instances, the chatbot does not inherently distinguish between factual or logical validity and the user's viewpoint. Instead, it primarily responds with text patterns predicted to be most consistent with the conversation's trajectory. Although some advanced models incorporate content filters or reference-checking systems to reduce factual inaccuracies, the inherent design of generative text models does not always prioritize challenging user assumptions. Some might argue that the chatbot's role is to serve user needs rather than to function as an adversarial fact-checker, but the risk is that it might inadvertently amplify misinformation or narrow a user's perspective by confirming preexisting beliefs (Chen et al., 2024).

Confirmation bias in generative AI chatbots might be exacerbated if the user consciously or unconsciously crafts prompts that narrow the scope of possible responses. For example, a user investigating a conspiracy theory could pose questions laden with presuppositions such as, "How did hidden organization Y secretly orchestrate event Z?" If the model has no directive to question these presuppositions, it might proceed to generate a narrative detailing how organization Y might have orchestrated event Z, thus reinforcing the user's belief in the conspiracy. This mechanism is reminiscent of leading questions in psychology and law, where the framing of the query strongly influences the nature of the response. Although humans can occasionally recognize the leading nature of such questions and respond skeptically, the generative AI chatbot is more likely to accept the premise and produce an accommodating narrative, unless it is explicitly designed to detect and challenge questionable premises.

In some cases, confirmation bias in chatbots might manifest more subtly. Rather than overtly reinforcing falsehoods, the chatbot might subtly omit relevant counterarguments or fail to present balanced perspectives. This omission could be rooted in the model's training distribution. If the training data includes a preponderance of text that aligns with a particular viewpoint, the system will be more likely to replicate that viewpoint in response to certain prompts. On the other hand, if balanced discussions of contentious issues are less prevalent in the training data, the system might have a diminished capacity to spontaneously introduce opposing viewpoints. The result could be a conversation that leans heavily in one direction, even if the user would have been receptive to a more varied treatment of the topic. This effect could be especially pronounced in domains where misinformation is more readily available or more heavily represented in the data, such as in certain fringe communities online.

The challenges presented by confirmation bias in generative AI chatbots may be particularly pronounced in high-stakes contexts. Individuals seeking medical or legal guidance, for instance, could inadvertently reinforce dangerous or illegal decisions if their queries presume the appropriateness of a particular course of action. The chatbot might comply with the premise that a certain therapy is effective or that a certain legal strategy is sound, without introducing the necessary caveats that a professional human expert would normally provide (Thirunavukarasu et al., 2023). Even in less critical contexts, the systematic reinforcement of user assumptions has the potential to undermine the user's ability to engage in more rigorous or reflective thinking. If the chatbot consistently accedes to the user's premises, it might become more difficult for users to encounter alternative views or to refine their understanding of complex issues.

While it is recognized that confirmation bias has a long history of discussion in social sciences, the notion that AI chatbots might replicate this bias in a computational manner is relatively new. The phenomenon has not been extensively measured or analyzed in the same way as other biases such as racial or gender-based biases (Chen et al., 2023). Yet the potential impact on user cognition, social discourse, and information ecosystems may be significant. The following sections delve into the specific mechanisms by which confirmation bias is embedded in generative AI architecture, as well as the associated risks and implications for both users and society at large.

## 4. Mechanisms of Confirmation Bias in Chatbot Architectures

The architecture of large language models is shaped by the goal of predicting linguistic patterns (Dütting et al., 2024). This generative capacity is often implemented through transformer-based architectures, which rely on attention mechanisms to process and generate text (Raiaan et al., 2024). In these architectures, the model learns relationships between tokens at multiple scales, capturing syntactic dependencies and semantic patterns from vast amounts of text data (Kumar et al., 2024). During inference, the model uses the prompt provided by the user to generate the most plausible continuation of the text. This structural foundation creates multiple pathways through which confirmation bias can materialize.

One potential mechanism involves the alignment of the model's probability distributions with user queries. When the user's prompt specifies or implies a particular stance, the model's response might be heavily influenced by that stance (Raschka, 2025). This effect arises because the model has effectively learned to treat the user's input as an important context that shapes subsequent token prediction (Ozdemir, 2025). The user's query, which might be laden with implicit or explicit assumptions, frames the generation space. The model, adhering to the principle of providing the most contextually relevant continuation, is inclined to produce output that maintains internal coherence with the prompt (Rothman, 2024, p. 3). If the prompt strongly suggests that a given viewpoint is correct, the model could reinforce that viewpoint in its output. This phenomenon is not necessarily malicious or reflective of an internal objective to confirm the user's biases; it is rather a direct product of the generative model's design, which privileges contextual consistency above all else.

A related mechanism stems from the iterative nature of user-chatbot interactions. Generative AI chatbots are typically configured to sustain multi-turn conversations, remembering previous user inputs and the bot's own responses throughout the dialogue (Alammar & Grootendorst, 2024). The history of the conversation effectively acts as an accumulating context that constrains future responses. If early in the conversation the user proposes a particular hypothesis, the chatbot's subsequent responses might increasingly assume that hypothesis to be valid, since consistency with conversation history is highly prioritized by many generative architectures (Rothman, 2024, p. 3). This iterative reinforcement can be thought of as a feedback loop: the user's initial question or statement sets the stage, the chatbot confirms it in an attempt to remain consistent, and further user queries delve deeper into that particular perspective, prompting further confirmations. Over multiple exchanges, it becomes progressively less likely that the chatbot will introduce contrary information that disrupts the established narrative, unless the user explicitly requests it.

Additionally, confirmation bias could be tied to the training corpus itself. During model pre-training, the system learns associations from textual data across many domains (Alammar & Grootendorst, 2024). If, in certain domains, the data is skewed toward specific viewpoints or if contradictory perspectives are underrepresented, the model's knowledge might similarly reflect that imbalance (Ozdemir, 2025). In such circumstances, when the user prompts a conversation within that domain, the model is more prone to produce responses reinforcing the perspective that it has seen most frequently in its training data, thereby unintentionally perpetuating the same skew. This phenomenon is not limited to any single issue or political viewpoint; it might arise in various domains where textual data distribution is uneven. Some of the largest and most influential language models have been trained on data sets that are not fully transparent or are only partially curated. This lack of transparency complicates efforts to understand how data distribution might influence the chatbot's tendencies toward confirmation or disconfirmation of user assumptions.

Moreover, the fine-tuning process introduced after pre-training can also influence how confirmation bias surfaces. Many generative AI chatbots undergo a stage known as instruction tuning or reinforcement learning from human feedback, aiming to make the model more aligned with user preferences or more readily applicable to real-world tasks (Alammar & Grootendorst, 2024). This alignment process often includes strategies to ensure the model remains polite, non-offensive, and user-centric. While such alignment is desirable for many reasons, there might be trade-offs. A system that is heavily optimized to satisfy user intent might become disinclined to challenge a user's premise. This risk could be exacerbated by the explicit instructions that developers provide to the chatbot, such as instructions stating that the chatbot should remain neutral or should not engage in contentious debates unless explicitly asked. Inadvertently, these instructions might discourage the chatbot from introducing contradictory evidence or alternative viewpoints.

Another factor that can reinforce confirmation bias relates to the user interface design and the user's mental model of the chatbot's capabilities. Some users might consider the chatbot an expert, or at least a credible source of information, and ask questions in a manner that presupposes a certain type of answer (Jeon & Lee,

2023). Users could also pick up on subtle signals from the chatbot that it is validating their viewpoint, encouraging them to pose follow-up questions that reinforce their pre-existing beliefs. The chatbot's capacity to recall context across multiple turns heightens the risk that the user will see the chatbot as corroborating evidence for their stance, when in fact the chatbot is merely fulfilling its generative task of producing coherent text consistent with the conversation's trajectory.

It must be noted that the presence of these mechanisms does not mean confirmation bias is guaranteed to appear in all chatbot interactions. There are many instances in which users ask straightforward factual questions where the chatbot's role is simply to retrieve or synthesize information in a neutral manner. However, the risk of confirmation bias becomes more pronounced when user inputs contain loaded assumptions or when the subject matter is controversial or poorly understood. In these scenarios, the model's design to maintain coherence, context, and user satisfaction may make it susceptible to systematically reinforcing the user's perspective. This issue underlines the importance of exploring the ethical implications and potential harms connected with confirmation bias in generative AI chatbots, topics addressed in the next section.

## 5. Risks and Ethical Implications

Confirmation bias in AI chatbots could potentially carry several risks, some of which might have wide-ranging ethical and societal impacts. These risks stem from the fact that generative chatbots, though highly sophisticated at producing human-like text, do not inherently possess the capacity to evaluate the truth-value or moral implications of the content they generate. They operate primarily on statistical associations rather than on reasoned judgments or introspective cognition. Consequently, when users engage with these chatbots and receive responses that confirm their preconceived notions, there may be a reinforcement of potentially harmful beliefs. In extreme cases, this can lead to the spread of misinformation, the intensification of conspiracy theories, and other problematic outcomes that could shape public discourse negatively.

One potential harm concerns the entrenchment of misinformation. In social media environments, echo chambers can form around sensational or controversial viewpoints, sometimes rooted in misleading or blatantly incorrect information (Nguyen, 2020). If chatbots systematically confirm the user's underlying assumptions—especially those assumptions that align with fringe ideologies or conspiracy theories—the user's confidence in those beliefs may grow. The chatbot's refusal (or inability) to introduce contradictory evidence can strengthen a false sense of validity for positions that lack empirical grounding. Over time, users who rely heavily on chatbots for knowledge acquisition might become more entrenched in erroneous or unbalanced views. This entrenched belief system, in turn, can create further social divisions, as individuals become less exposed to disconfirming information.

Furthermore, there are ethical questions about the chatbot's responsibility to guide users toward accurate or balanced information (Pressman et al., 2024). While human experts, such as teachers or doctors, might detect misguided assumptions and provide corrective feedback, generative AI chatbots often respond without critical scrutiny. The user might interpret the chatbot's detailed answer as expert validation. The lack of an explicit safeguard against perpetuating one-sided narratives raises concerns about whether developers and providers of such technology might bear some ethical responsibility for the outcomes of the system's interactions. Regulators and policymakers might eventually demand that chatbot systems incorporate design features that minimize confirmation bias, prompting new debates about the boundaries of free inquiry, user autonomy, and paternalistic algorithmic design.

Beyond misinformation, confirmation bias in generative AI chatbots can pose risks for individual decision-making. Individuals sometimes consult chatbots for advice on significant personal or professional choices, such as financial investments or career planning (Duan et al., 2019). If the chatbot systematically validates the user's initial inclination, it might fail to present alternative approaches that could be more advantageous.

This could harm the user's interests, especially if the user uncritically acts on the chatbot's recommendations. It is possible that chatbot developers could claim they bear no legal liability for incorrect guidance, but ethical questions remain regarding whether it is appropriate to deploy a system that might inadvertently guide users away from well-informed decisions.

The risk of social fragmentation is another potential outcome. If users engage with chatbots as interactive information sources, they might find that the chatbot's responses consistently resonate with their existing worldviews (Xue et al., 2024). This phenomenon could be exacerbated by recommendation systems on social media platforms that already steer users toward content aligning with their preferences. Together, these systems might create a self-reinforcing loop in which AI-driven interactions fail to challenge users' worldviews, further narrowing the scope of dialogue and common understanding (Chen et al., 2023). Over time, this could contribute to increasing polarization in societal and political contexts, as groups become less likely to encounter perspectives that differ from their own. Although AI chatbots alone may not be the primary cause of polarization, the extent to which they encourage confirmation bias could nonetheless be significant.

Moreover, there may be legal ramifications. In certain regulated domains—such as medicine, finance, or legal services—chatbots that confirm a user's biases could inadvertently give rise to negligent or harmful outcomes (Nazi & Peng, 2024). If a user inquires about self-medication strategies, for instance, and the chatbot's response reinforces the user's original premise, it might result in unsafe behavior. Some jurisdictions have enacted or are considering regulations that establish standards for the deployment of AI in high-stakes areas (Minssen et al., 2023). If chatbots fail to challenge potentially dangerous assumptions, developers might be scrutinized for insufficient oversight of the system's outputs, especially when real harm can be traced to biased or misleading responses.

Concerns about user autonomy are also relevant. Confirmation bias in a chatbot does not merely reflect an existing user belief; it can also shape future beliefs. Users may not be fully aware that the chatbot is not designed to question the assumptions in their prompts. If a system subtly steers the user away from balanced consideration or critical thinking, the user may become effectively less autonomous in formulating informed opinions. Questions could be asked about whether AI design choices that prioritize user satisfaction inadvertently undermine the user's ability to engage with diversity of thought. In that sense, the tension between user-centered design and the need for balanced, introspective dialogue becomes more pronounced.

Despite these risks, it is worthwhile to consider that confirmation bias in AI chatbots might sometimes produce neutral or even beneficial outcomes, especially if the user's initial assumptions are factually accurate or benign. The chatbot's affirmation might reinforce correct knowledge or positive behaviors. Nonetheless, from an ethical and societal standpoint, the capacity to inadvertently reinforce problematic or harmful beliefs is a concern that merits careful attention. Recognizing these risks underscores the importance of developing and adopting mitigation strategies that can reduce the incidence and impact of confirmation bias. The next section addresses these strategies, exploring how technical, design, and policy interventions might help ensure that generative AI chatbots serve as more balanced and responsible tools.

## 6. Mitigation Strategies
Efforts to mitigate confirmation bias in generative AI chatbots may take several forms, encompassing improvements in model design, user interface modifications, and policy or regulatory interventions. Although none of these approaches alone may fully eliminate the possibility of bias, a combination could prove more effective. Given that this is still a nascent area of research, developers, researchers, and policymakers might benefit from a variety of techniques aimed at making AI-generated conversations more balanced and reflective.

One approach could involve training the chatbot to detect and flag presuppositions or highly loaded assumptions in user queries. Techniques from natural language understanding might be adapted to recognize signals that a user is operating within a narrow viewpoint or searching for confirmation of a particular belief. The chatbot could then offer a prompt indicating that there may be alternate ways of interpreting the issue. For example, in response to a query such as, "Explain why group A is responsible for problem B," the chatbot might respond, "It appears this question presupposes that group A is indeed responsible. Would you like to explore different perspectives or evidence that counters this view?" This function would not necessarily force the user to consider alternate viewpoints, but it might gently nudge users to reflect on their assumptions, thereby introducing some friction against the automatic acceptance of a single narrative.

In some cases, developers may consider building specific "contradiction modules" into the chatbot. These modules might be designed to generate or retrieve arguments that contest the user's prompt, possibly referencing authoritative sources or recognized counterarguments. The chatbot could ask if the user would like to see opposing views or data. Such a mechanism might be optional, ensuring that user autonomy is not undermined. However, it could address the issue that many users, particularly those deeply invested in certain beliefs, rarely seek disconfirming evidence. By making disconfirming evidence more readily accessible, the chatbot might reduce the risk of reinforcing biases. The challenge is to implement this strategy in a manner that respects user preferences and does not appear overly paternalistic.

User interface design might also be leveraged. For instance, presenting multiple plausible answers simultaneously could encourage users to compare and contrast different lines of reasoning. If the interface offers multiple responses, each representing a distinct viewpoint, the user can see at a glance that alternate interpretations exist. This approach may be akin to displaying various search results, some confirming the user's premise and others contradicting it. Although it would require more computational resources and a thoughtful design, it has the potential to mitigate the single-minded reinforcement of user assumptions.

Another angle is the use of metadata and transparency. Chatbots could identify which sources or segments of the training data they are drawing from. If a user asks about a contentious topic, the chatbot might indicate that it has found a variety of perspectives in its training corpus, referencing specific disclaimers or reliability ratings for each. This transparency could help users evaluate the credibility of the chatbot's responses and might also make them more aware of the potential for confirmation bias (Afroogh et al., 2024). Although this approach could become unwieldy, especially given the massive size of modern language model training sets, partial measures that reveal the chatbot's evidence or rationale might be beneficial.

On a more policy-oriented level, guidelines or regulations might mandate that generative AI systems incorporate certain features to reduce bias reinforcement. Such policies might require AI developers to integrate content moderation or disclaimers that warn users about potential biases or the limitations of AI-generated responses. Policymakers could also encourage or require that training datasets be more systematically balanced, especially in high-stakes domains. Furthermore, auditing frameworks could be developed to assess whether chatbots systematically display confirmation bias under controlled testing conditions. These frameworks might feed into certification or rating systems for AI systems, allowing consumers to make more informed choices about which chatbots they interact with.

Another promising area lies in the potential for dynamic or interactive fine-tuning. Rather than relying solely on a static model, developers might incorporate user feedback loops in which signals that the model is consistently supporting a controversial or fringe assumption trigger retraining or rebalancing operations. The challenge is that such feedback loops have the potential to be gamed by coordinated efforts that push the model in a particular direction. Care must be taken to ensure that the feedback mechanism itself does not become a tool for cementing biases. Nonetheless, a well-structured system of iterative improvement could help generative chatbots learn to introduce a broader range of perspectives when confronted with prompts that seem to desire confirmation.

Ultimately, success in mitigating confirmation bias could depend on user education as much as on technical interventions. Informing users about the inherent limitations of generative AI, and reminding them that the model does not possess an inherent ability to verify the veracity of its statements, might encourage more cautious consumption of chatbot outputs. Users could be guided to treat the chatbot's responses as starting points for inquiry rather than definitive conclusions. To some extent, this educational angle aligns with broader media literacy efforts that stress the importance of critical thinking when reading any text—be it produced by humans or AI. As generative AI chatbots become ubiquitous, user literacy in understanding the nature and constraints of these models may be vital in preventing the uncritical acceptance of bias-reinforcing content.

These mitigation strategies hold promise, but it remains uncertain to what degree they will address the core challenge of confirmation bias. The next section turns to the topic of future research directions, proposing areas of empirical study and methodological innovation that might be necessary to more precisely characterize and counteract the mechanisms through which confirmation bias arises in generative AI chatbots.

### 7. Future Research Directions

Confirmation bias in generative AI chatbots is a relatively new domain of inquiry, and substantial research is likely required to develop a thorough understanding of the phenomenon and effective mitigation strategies. Future work could begin with systematic, empirical investigations into how these biases manifest across different user populations, domains, and model types. Large-scale user studies might examine whether individuals with strong predispositions on various topics—such as politics, health, or conspiracy theories—are indeed reinforced in their beliefs when interacting with generative AI chatbots. These studies could measure shifts in user perspectives over time, comparing interactions with AI systems that contain various bias-mitigation features to those that do not.

On the technical front, more sophisticated techniques for detecting and countering bias-laden prompts or dialogues could be explored. Research might focus on developing specialized classifiers that can identify confirmation-seeking behavior in user inputs. Collaborative efforts between cognitive psychologists and AI developers might yield insights into how these classifiers could be best trained and evaluated. There is also a need for deeper exploration of the interpretability of large language models, to better grasp the internal mechanics that lead to confirmation bias-like outputs. Although interpretability methods such as attention visualization have been introduced, they often fall short of offering a comprehensive explanation of generative model decisions.

Researchers may also investigate how the fine-tuning process, or reinforcement learning from human feedback, could be structured to reduce confirmation bias. This might entail designing new forms of user feedback that specifically ask the model to present alternative perspectives. Additionally, reevaluating the trade-offs between user satisfaction and critical perspective could be important, as user-focused metrics may inadvertently promote excessive agreement with user assumptions. Studies could focus on measuring user satisfaction in scenarios where chatbots challenge or question user viewpoints, determining whether such challenges are broadly accepted or trigger user dissatisfaction and disengagement.

From a social science perspective, longitudinal studies of chatbots' role in shaping online discourse might reveal how confirmation bias in these systems interacts with existing tendencies in social media. It may be prudent to consider the broader information ecosystem in which these chatbots operate. For instance, many users might simultaneously use AI-generated text in social media discussions or consult chatbots while reading articles and blog posts. The interplay between AI-generated confirmations and human interactions could be complex, potentially magnifying echo chambers. Understanding and modeling these dynamics would demand interdisciplinary collaboration among computer scientists, sociologists, communication experts, and policymakers.

Cross-cultural considerations also deserve attention. Training data for language models often stems heavily from English-language sources and from certain regions. If confirmation bias is partly a function of training data distributions, then the manifestation of bias may differ significantly when the chatbot is deployed in other languages or cultural contexts. Research is needed to determine how well existing solutions for mitigating confirmation bias generalize to various linguistic or cultural settings. Additionally, the ethical standards for challenging user assumptions might differ across societies, necessitating culturally sensitive guidelines for chatbot design.

Finally, the advancement of AI governance frameworks that address generative models could help shape future research directions. Regulatory agencies may request audits of chatbot outputs to measure different forms of bias, including confirmation bias. Encouraging or requiring transparency about model training data or model performance metrics could incentivize developers to investigate and mitigate biases thoroughly (Walmsley, 2021). This top-down approach might act in tandem with grassroots or community-based initiatives that track and report bias-related concerns. The open-source AI community, for example, might collectively experiment with different approaches to data curation, model architectures, or user interface designs, pooling insights on which methods prove most effective in mitigating confirmation bias (Schmidt et al., 2020). Through these avenues of inquiry, it is possible that the field will arrive at a more profound understanding of how generative AI chatbots can be designed and deployed in ways that minimize the unintentional reinforcement of user assumptions.

## 8. Conclusion

Confirmation bias is recognized in cognitive science as a potent force shaping human thought and discourse. In generative AI chatbots, this bias may be replicated in subtle but potentially consequential ways. Because these models are built to produce text that aligns with user prompts, they may inadvertently reinforce assumptions rather than challenge them. Across contexts, from casual question answering to more sensitive applications such as health or financial advice, the systematic confirmation of user expectations could influence decision-making, public discourse, and the broader information ecosystem.

While many of the mechanisms behind confirmation bias in generative AI chatbots remain to be fully elucidated, researchers have begun to identify possible pathways, including the alignment of probabilistic text generation with user prompts, conversation history dependencies, imbalances in training data, and alignment methods that prioritize user satisfaction. Observers have also pointed out multiple risk factors, such as the propagation of misinformation, increased polarization, and compromised user autonomy.

Mitigation strategies may involve technical interventions, like detecting loaded assumptions in user prompts, offering multiple perspectives, and embedding contradiction modules. User interface design and policy-level initiatives could further contribute to reducing the likelihood of biased reinforcement. Still, many open questions remain regarding the optimal balance between user guidance and user freedom, the extent to which these solutions generalize across different cultures and linguistic contexts, and how to evaluate chatbot performance in terms of bias prevention.

Further research is needed to better define the contours of confirmation bias in AI-generated dialogue, develop robust measurement techniques, and test interventions at scale. This line of inquiry is likely to be pursued in tandem with broader efforts to ensure the transparency, accountability, and trustworthiness of generative AI systems. By engaging in rigorous investigation, interdisciplinary collaboration, and thoughtful policy development, it is possible to move toward AI chatbots that do not merely confirm what users believe but also empower them to explore the full complexity of the world around them.